\documentclass[reprint,
superscriptaddress,
amsmath,amssymb,
aps,
prl,
longbibliography,
showpacs
]{revtex4-1}
\usepackage{graphicx}

\usepackage{color}

\begin{document}

\title{Negative Wigner function at telecommunication wavelength from homodyne detection}
\author{Christoph~Baune}
\affiliation{Institut f\"ur Laserphysik und Zentrum f\"ur Optische Quantentechnologien, Universit\"at Hamburg, Luruper Chaussee 149, 22761 Hamburg, Germany}
\affiliation{Institut f\"ur Gravitationsphysik, Leibniz Universit\"at Hannover and Max-Planck-Institut f\"ur Gravitationsphysik (Albert-Einstein-Institut), Callinstrasse 38, 30167 Hannover, Germany}
\author{Jarom\'ir~Fiur\'a\v{s}ek}
\affiliation{Department of Optics, Palack\'y University, 17. listopadu 12, 77146 Olomouc, Czech Republic}
\author{Roman~Schnabel}
\email{roman.schnabel@physnet.uni-hamburg.de}
\affiliation{Institut f\"ur Laserphysik und Zentrum f\"ur Optische Quantentechnologien, Universit\"at Hamburg, Luruper Chaussee 149, 22761 Hamburg, Germany}

\date{\today}

\begin{abstract}
Quantum states of light having a Wigner function with negative values represent a key resource in quantum communication and quantum information processing.
Here, we present the generation of such a state at the telecommunication wavelength of 1550\,nm. The state is generated by means of photon subtraction from a weakly squeezed vacuum state and is heralded by the `click' of a single photon counter. Balanced homodyne detection is applied to reconstruct the Wigner function, also yielding the state's photon number distribution. The heralding photons are frequency up-converted to 532\,nm to allow for the use of a room-temperature (silicon) avalanche photo diode. The Wigner function reads $W(0,0)=-0.063\pm0.004$ at the origin of phase space, which certifies negativity with more than 15 standard deviations.
\end{abstract}

%\keywords{Quantum Physics, Quantum Information, Photonics}
\maketitle

\emph{Introduction} -- 
A quantum state having a negative Wigner function is distinct from the class of semi-classical or `classical' states.
The Gottesman-Knill theorem states that a wide class of systems can be efficiently simulated with classical methods \cite{Gottesman1999, Bartlett2002}. 
This is not true for systems showing negative Wigner functions.
While quantum computers can solve certain problems much faster than ordinary computers, a quantum computer outperforms its classical counterpart only when states with a negative Wigner function are used \cite{Mari2012}.

The Wigner function \cite{Wigner1932} provides the full information about a quantum state. It is a quasi-probability function of the phase space spanned by the quadrature amplitudes $\hat{X}$ and $\hat{Y}$. These observables obey the commutation relation $[\hat{X},\hat{Y}] = i\,$ and cannot be measured simultaneously with arbitrary precision. Measurements on an ensemble, however, 
allow for a `tomographic' approach in which marginal distributions are measured, from which the Wigner function can be reconstructed \cite{Vogel1989,Lvovsky2009}. The marginal distributions are measured with a balanced homodyne detector (BHD).

Quantum states with (partly) negative Wigner functions can efficiently be generated from squeezed vacuum states by conditioning on successful photon subtraction, or from two-mode vacuum states by conditioning on the successful detection of a photon in the idler mode. 
Squeezed vacuum states of perfect purity show either zero or an even number of photons. For \emph{weakly} squeezed, pure vacuum states the probability of measuring more than two photons can be neglected, and successful photon subtraction heralds in very good approximation a single photon Fock state \cite{Dakna1997,Lvovsky2001,Molmer2006}, which shows the strongest negativity of all quantum states in the Wigner representation of $-1/\pi$. 

Previous works demonstrated negative Wigner functions with balanced homodyne detection in both pulsed and continuous-wave optical setups at wavelengths around 800 nm and 1064 nm \cite{Lvovsky2001,Zavatta04,Ourjoumtsev06,Neergaard-Nielsen06,Neergaard-Nielsen2007a,Wakui07,Gerrits10,Huang15}. 
A negative Wigner function at the prominent telecommunication wavelength of $1550$\,nm, however, could not be observed (with homodyne detection) so far. 
Photon subtraction at telecom wavelength was implemented with the use of superconducting transition-edge sensor \cite{Namekata10}, but negative values of the Wigner function were not achieved due to low modal purity. 
A negative Wigner function of conditionally generated single photon states at telecom wavelength was inferred by measurements with single photon detectors \cite{Harder16}. 

Here we reconstruct for the first time, to the best of our knowledge, a negative Wigner function of a quantum state at the wavelength of 1550\,nm from balanced homodyne detector data. Our approach follows that of earlier works at other wavelengths, i.e.~successful photon subtraction heralds the generation of the state.  As a specific feature of this work, the detection of the heralding photon is assisted by on-the-fly frequency up-conversion to 532\,nm to allow for the exploitation of low noise, high quantum efficiency silicon avalanche photo detectors operated at room temperature.

\begin{figure}[bt]
	\centering
	\vspace{0mm}
	\includegraphics[width=0.46\textwidth]{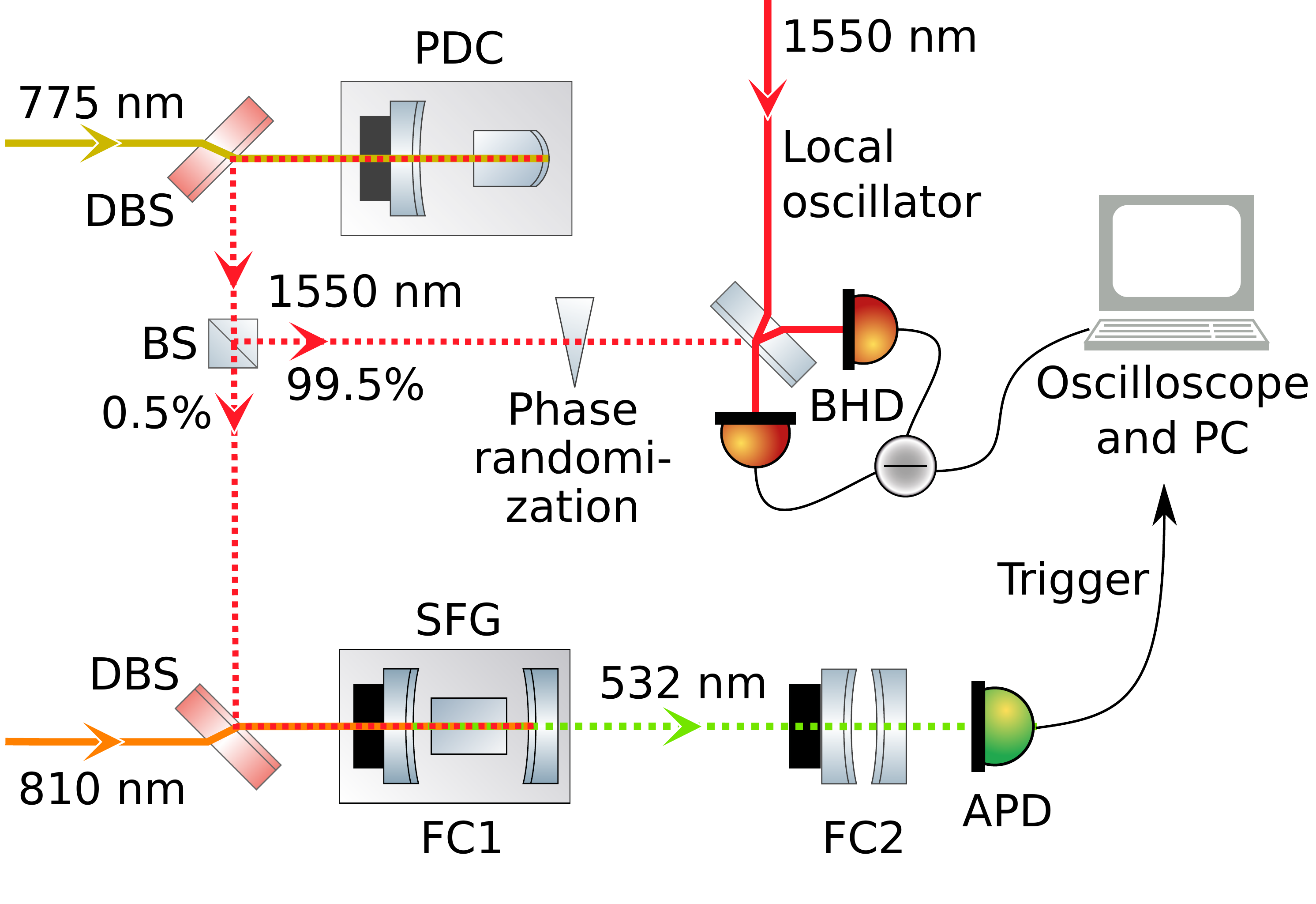}
		\vspace{-2mm}
	\caption{(Color online) Experimental setup for quantum state tomography on photon-subtracted states at the telecommunication wavelength of 1550\,nm \cite{Lvovsky2009}. First, squeezed vacuum states of 1550\,nm light are produced via cavity-enhanced parametric down-conversion (PDC). This process requires a pump field at 775\,nm. For pump intensities far below the oscillation threshold (about 1\% of about 80\,mW) 
	%\tcr{( What is the probability of 2 photons compared to 4 in this case? )}
	weakly squeezed vacuum states are produced. A small part of the squeezed state (0.5\%) is tapped off at a beam splitter (BS) and is filtered and frequency up-converted to 532\,nm in a sum-frequency generation (SFG) cavity, pumped with about 140\,mW at 810\,nm. The up-converted mode at 532\,nm passes a second filter cavity (FC2) and is detected by a silicon avalanche photo detector (APD). A detection event of the APD projects the remaining state at 1550\,nm into a photon-subtracted state that, for weak squeeze factors, has a high overlap with a single photon Fock state. The photon-subtracted state is phase-randomized and analyzed via optical quantum state tomography using a balanced homodyne detector (BHD) \cite{Lvovsky2001}. DBS: dichroic beam splitter; $\blacksquare\!\!\blacksquare\!\!\blacksquare$: piezo-electrical positioner for keeping cavities on resonance.}
	\label{fig:Setup}
\end{figure}

\emph{Experiment} -- 
The experimental setup for quantum state tomo\-graphy on photon-subtracted states is shown in Fig.\,\ref{fig:Setup}. 
The ensemble of identical squeezed vacuum states of light at 1550\,nm is generated by cavity-enhanced parametric down-conversion (PDC) in periodically poled KTiOPO$_4$ (PPKTP) \cite{Mehmet2011}. A small fraction of the squeezed states (0.5\%) is tapped by a beam splitter, filtered and finally detected by an avalanche photo diode (APD). The remaining part of the squeezed states is analyzed via balanced homodyne detection (BHD). 
Every click of the APD triggers an oscilloscope (\textsc{Agilent} \textit{DSO 7014B}), which records the BHD signal for a short time period that contains the information about the single-photon subtracted state. 
The oscilloscope has a sampling rate of 2\,GSa/s and is controlled with a python program via the VISA protocol. 
The recorded raw data of every segment are directly transferred to a PC for precise extraction of the state's information.
	
The filtering via FC1 and FC2 is required to match the temporal and spectral mode that are detected by APD and BHD \cite{Kumar2012}. The PDC cavity has a linewidth of 120\,MHz (FWHM) and a free spectral range (FSR) of 3.6\,GHz.
The BHD is able to record the complete spectrum of a single longitudinal cavity mode since it has a flat response and a dark noise clearance of more than 15\,dB up to 200\,MHz with a local oscillator power of 15\,mW, disturbed just by a few electronic pick-up peaks in the dark noise of unknown origin above 100\,MHz. The PDC cavity, however, also produces photon pairs in longitudinal modes separated by multiples of twice the cavity's free spectral range of several gigahertz. These spectral components must not arrive at the trigger APD since the BHD is not able to measure them.

The filtering of the trigger mode is performed by two subsequent transmissions through cavities having free spectral ranges that differ from each other and also from that of the PDC cavity to achieve efficient filtering. 
The first cavity has an FSR of about 2.35\,GHz. It is not an empty cavity but contains an optically pump nonlinear medium and up-converts the trigger mode from 1550\,nm to 532\,nm using sum-frequency generation (SFG).  
The up-conversion efficiency is up to $(90.2 \pm 1.5)\,\%$ \cite{Baune2014}. 
The `SFG cavity' is doubly resonant for the intense pump wavelength at 810\,nm and the initial trigger mode wavelength at 1550\,nm. Ideally, double resonance is achieved simultaneously with optimum quasi-phase matching of the nonlinear medium (PPKTP) for all three wavelengths involved. In practice, the quasi-phase matching is not optimal when reaching double resonance, which reduces the nonlinearity. This, however, can be compensated for by increasing the pump field intensity; and even has  an advantage. Due to the small phase mismatch, the free spectral ranges of the cavity at the two wavelengths differ and neighboring free spectral ranges are not simultaneously resonant. This effect improves the filtering strength for photons from higher order longitudinal modes of the squeezing resonator. 

The frequency up-conversion of the trigger field from 1550 to 532\,nm is mainly done to enable the use of commercially available, room temperature high-quantum efficiency low-noise and easy-to-use silicon APDs, which are not responsive for infrared wavelengths.
APDs for 1550\,nm based on InGaAs chips or superconducting sensors require cooling, in the latter case even down to cryogenic temperatures, to achieve comparable detection efficiencies and noise performance \cite{Hadfield2009}.
Details on the up-conversion setup can also be found in \cite{Baune2014,Fiurasek2015,Baune2016}. 
The second filter stage was implemented with a short linear filter cavity (FC2).
This cavity consists of two half-inch mirrors with a nominal reflectivity of 99\% and a spacing of 3.2\,mm yielding a linewidth (FWHM) of 150\,MHz and a free spectral range of 47\,GHz. 

The lengths of the filter cavities are adjustable via piezo-electric elements to which one of the cavity mirrors was attached. 
The length of FC1 was electro-optically controlled using light at 810\,nm, and its length stabilized on resonance. FC2 was intrinsically stable and was manually set on resonance.

\emph{Data taking and analysis} --
In a first step, $10^4$ triggered segments are recorded with a blocked signal port of the BHD to obtain a reference in terms of vacuum states. Each recorded segment consists of $10^3$ data points representing 250\,ns of data before and after each trigger event. In a second step, the signal port is opened and $K=5\cdot10^4$ triggered segments are recorded.

Initially it is not known which points of the segments (at time $\tau$) contain the information about the photon-subtracted state, and to what extend. 
These informations are represented by the (temporal) mode function $f_m(\tau)$. To determine $f_m(\tau)$ experimentally, 
one makes use of the fact, that the variance of an arbitrary quadrature amplitude of a Fock state $|n\rangle$ is higher than that of the vacuum state $|0\rangle$ (by the factor $(2n+1)$). When comparing the segments with each other, points that contain information about the photon-subtracted state thus show an increased variance $V(\tau)$ with
\begin{equation}
V(\tau)=\kappa^2 |f_m(\tau)|^2+V_{\mathrm{0}} \, .
\label{eq:1}
\end{equation}
Assuming $f_m(\tau)$ is real and positive, we get $f_m(\tau)=\sqrt{V(\tau)-V_{\mathrm{0}}}/\kappa$, where $\kappa$ is a normalization factor and $V_0$ is an asymptotic value of $V(\tau)$ for large distances from the mode function's peak \cite{Morin2013}. 

\begin{figure}[h]
	\begin{minipage}{0.48\textwidth}
	\centering \includegraphics[width=0.75\textwidth]{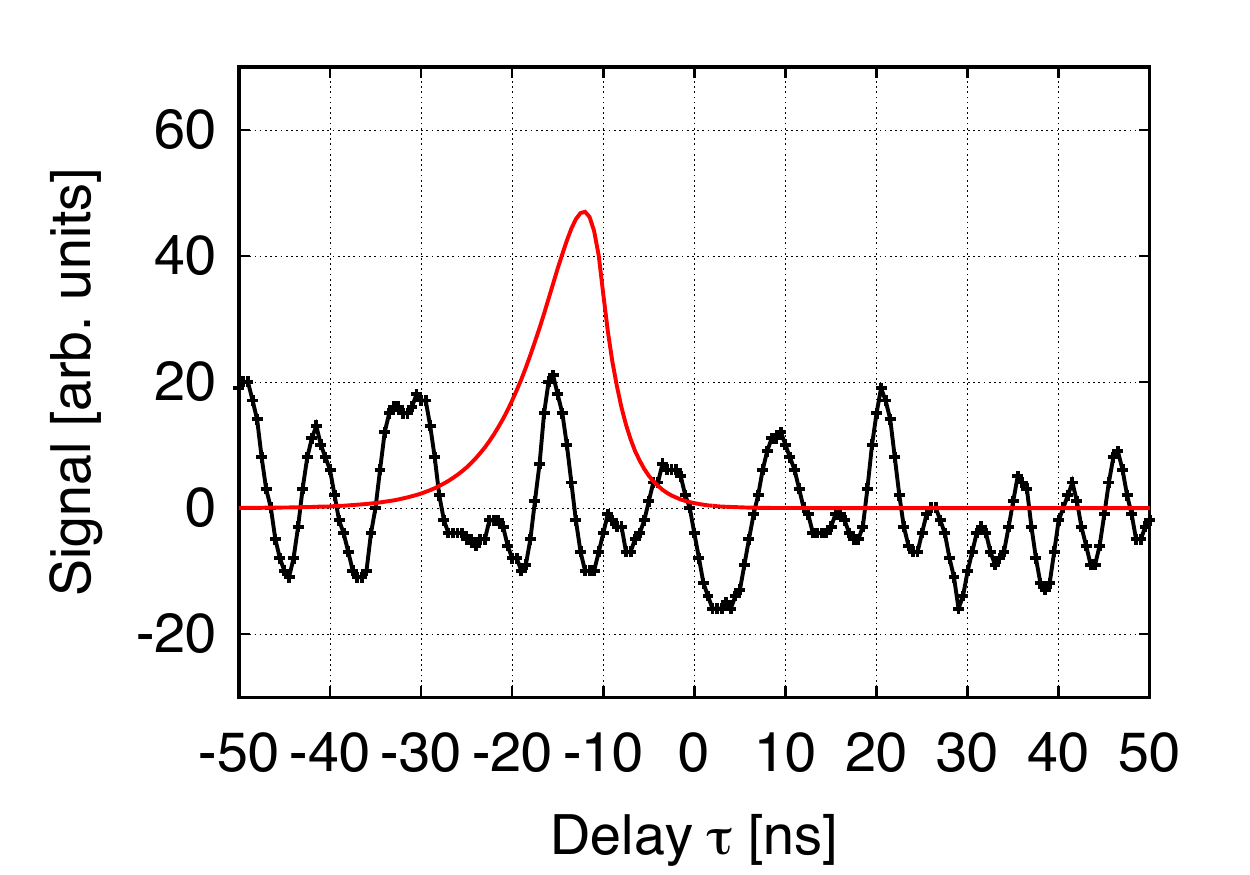}
	\end{minipage}
		\hspace{2mm}%\vline\hspace{1mm}
	\begin{minipage}{0.48\textwidth}
	\centering	\includegraphics[width=0.75\textwidth]{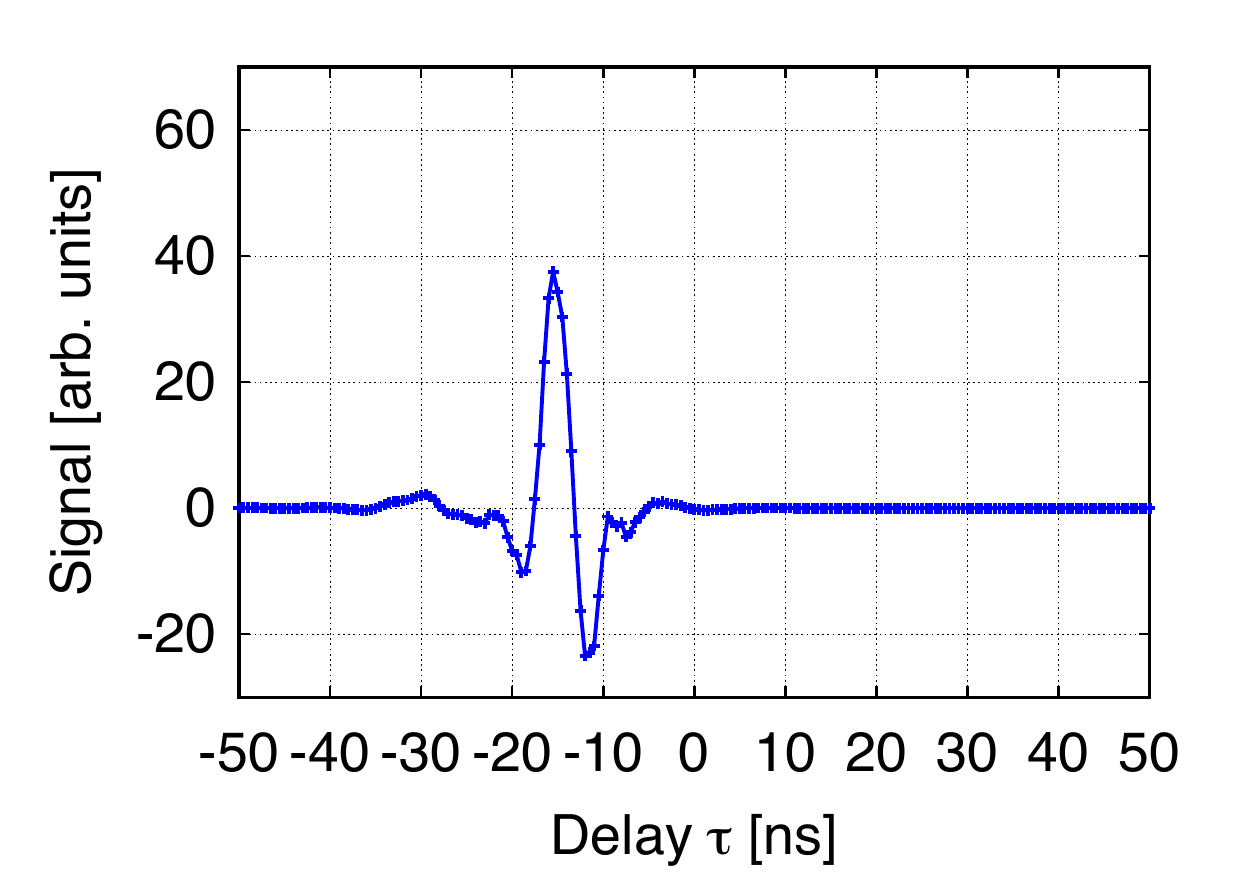}	
	\end{minipage}
	\caption{\textit{Top:} Single segment of the BHD data as recorded by the oscilloscope together with the mode function $f_m(\tau)$ (smooth curve) as derived from the variance of all segments for individual times around the APD trigger events, see Eq.\,(\ref{eq:1}).
	\textit{Bottom:} Single segment  multiplied with $f_m(\tau)$. Its integral provides a single quadrature amplitude value $X$ of the phase-randomized photon-subtracted state.
	}
	\label{fig:Segment}
\end{figure}

Fig.\,\ref{fig:Segment} (top) shows a recorded example segment, including the mode function of our setup as derived from all segments. The peak of  $f_m(\tau)$ is not positioned exactly at the trigger time (i.e. $\tau=0$) but slightly earlier due to optical delays and different electronic response times of the APD and the BHD, cf.~the discussion in references \cite{Baune2014, Fiurasek2015}. 

In the next step of the data analysis the quadrature amplitudes of the photon-subtracted states are obtained. Every segment results in one quadrature value. The segments are post-processed 
by multiplying the mode function $f_{\mathrm{m}}(\tau)$ to the raw data.
Fig.\,\ref{fig:Segment} (bottom) shows an example for one segment. 
The quadrature value is obtained by integrating the filtered segment data. The vacuum reference data serve for calibration of the vacuum noise level, and the measured quadrature amplitudes are normalized such that their variance yields $\Delta^2 \hat X_{\mathrm{vac}} = 1/2$.

\begin{figure}[htb]
\centering
	\includegraphics[width=8.9cm]{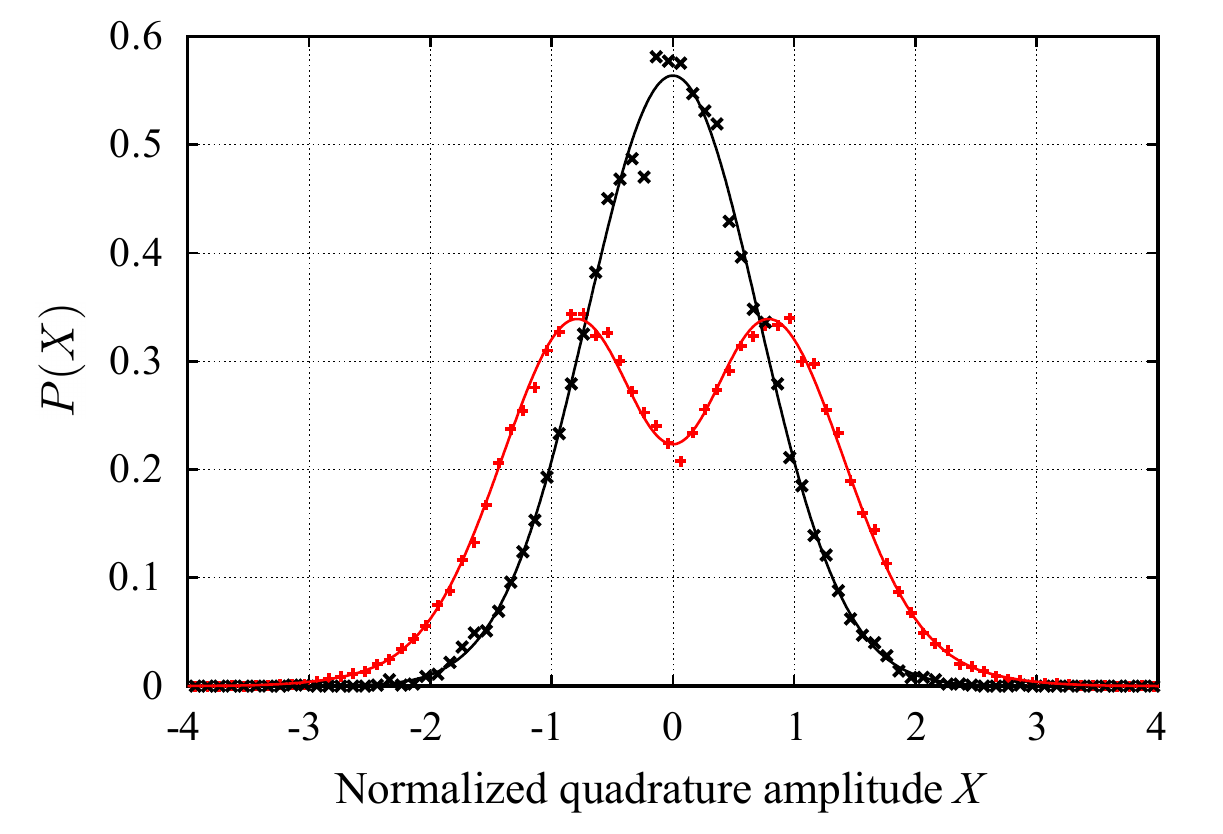} \vspace{-4mm}
\caption{Normalized marginal probability distributions of measured quadrature amplitude values. Solid lines represent theoretical models. The singly peaked curve corresponds to the vacuum state, i.e. a Gaussian distribution with variance $\Delta^2 \hat X_{\mathrm{vac}} = 1/2$. The doubly peaked curve corresponds to the single-photon subtracted state. The dip of the probability around the origin is clearly visible and characteristic for a single photon state. The solid curve is a model for a state with $p_0=0.39$, $p_1=0.57$, $p_2=0$, $p_3=0.03$, $p_4=0$, $p_5=0.01$ where $p_n$ is the probability of Fock state $|n\rangle$.}
\label{fig:QuadHistogram}
	\end{figure}

\emph{Results} -- 
The measured quadrature values for the vacuum state and the photon-subtracted squeezed state are shown in histograms in Fig.\,\ref{fig:QuadHistogram}, scaled to obtain normalized probability distributions. 
The solid lines represent the theoretical Gaussian quadrature distribution of a vacuum state $\pi^{-1/2} e^{-X^2}$ and the fitted model for a mixture of Fock states $|n\rangle$ with $n$ from zero to five.

Since the phase between the BHD local oscillator and the signal input was not stabilized but freely drifting, the homodyne detection samples the phase averaged quadrature distribution 
$P(X_{\overline \vartheta})= \frac{1}{2\pi}\int_0^{2\pi}P(X_\vartheta) d\vartheta,$ \cite{Munroe95}, where $\hat X_\vartheta = \hat X \rm{cos}\vartheta + \hat Y \rm{sin}\vartheta$ denotes the rotated quadrature and $\vartheta$ is the random phase.  
This is equivalent to measuring a quadrature distribution of a phase-randomized state, whose density matrix is diagonal in Fock basis, $\tilde{\rho}=\sum_{n} p_n|n\rangle\langle n|$. The quadrature distributions $P(X) = P(Y) = P(X_{\vartheta})$ of such a state  depend only on the photon number distribution $p_n$,
	\begin{equation}
	P(X) =\sum_{n=0}^\infty p_n Q_{n}(X),
	\label{Pqformula}
	\end{equation}
where 
\begin{equation}
Q_n(X)= \frac{1}{\pi^{1/2}2^n n!} H_n^2(X) e^{-X^2}
\end{equation}
denotes the quadrature distribution of Fock state $|n\rangle$. For practical calculations, a cut-off $N$ in the Fock state expansion needs to be introduced in Eq.~(\ref{Pqformula}).
A fit to the data with $N=5$ yields the probabilities $p_0=0.39$, $p_1=0.57$, $p_2=0$, $p_3=0.03$, $p_4=0$, $p_5=0.01$.

Additionally, we applied a maximum likelihood estimation algorithm to reconstruct the photon number distribution $p_n$. The likelihood function can be expressed as
$\mathcal{L}=\prod_{k=1}^K P(X_{k})$, where $K$ denotes the total number of detected quadrature values ($5\cdot 10^4$). 
The probability $p_n$ that maximizes  $\mathcal{L}$ can be found by iterative expectation-maximization algorithm \cite{Dempster77,Vardi93,Rehacek01}, whose single iteration reads	
\begin{equation}
p_m^{(j+1)}=\frac{p_m^{(j)}}{K} \sum_{k=1}^K \frac{Q_m(X_{k})}{\sum_{n=0}^{N} p_n^{(j)} Q_n(X_{k})},
\end{equation}
and a uniform distribution $p_{n}=1/(N+1)$ is chosen as the starting point of the iterations. As a result the following estimates of photon number probabilities are obtained:
$p_0=0.392,$ $p_1=0.572,$ $p_2=0.003$, $p_3=0.028,$ $p_4=0.004,$ $p_5=0.001$.
These very well agree with the values already obtained from a fit of the quadrature distribution, cf. Fig.~\ref{fig:QuadHistogram}.
The relatively high vacuum contribution $p_0=0.392$ is caused by false trigger events, which originate from dark counts of the APD (10\%) and imperfect filtering of higher-order longitudinal mode photon pairs from the PDC resonator and photons at 810\,nm (25\%), and optical losses in the path to the BHD.
The single-photon subtracted state is detected with an overall efficiency of about 90\%, which includes propagation losses, limited detection efficiency of the homodyne detector and the beam splitter to tap off a small part for the trigger path. 
This efficiency is inferred with an auxiliary squeezing/anti-squeezing measurement when the PDC cavity is operated with much higher parametric gain, in analogy to the procedure used in \cite{Vahlbruch2016}. 
The imperfections mentioned above explain the overall probability of about 40\% for detecting the vacuum state ($p_0 = 0.392$). In principle, the vacuum contribution can be considerably reduced (and thus that of a single photon considerably increased)  by adding another filter cavity in front of the APD and by improving the detection efficiency of the BHD. The highest probabilities of single photons in a state analysed by balanced homodyne detection so far ($p_1 \approx 0.79$) were achieved in \cite{Morin2013,Ogawa2016} at the wavelengths of 1064\,nm and 860\,nm, respectively.

\begin{figure}[hbt] 
\centering
\includegraphics[width=8.8cm]{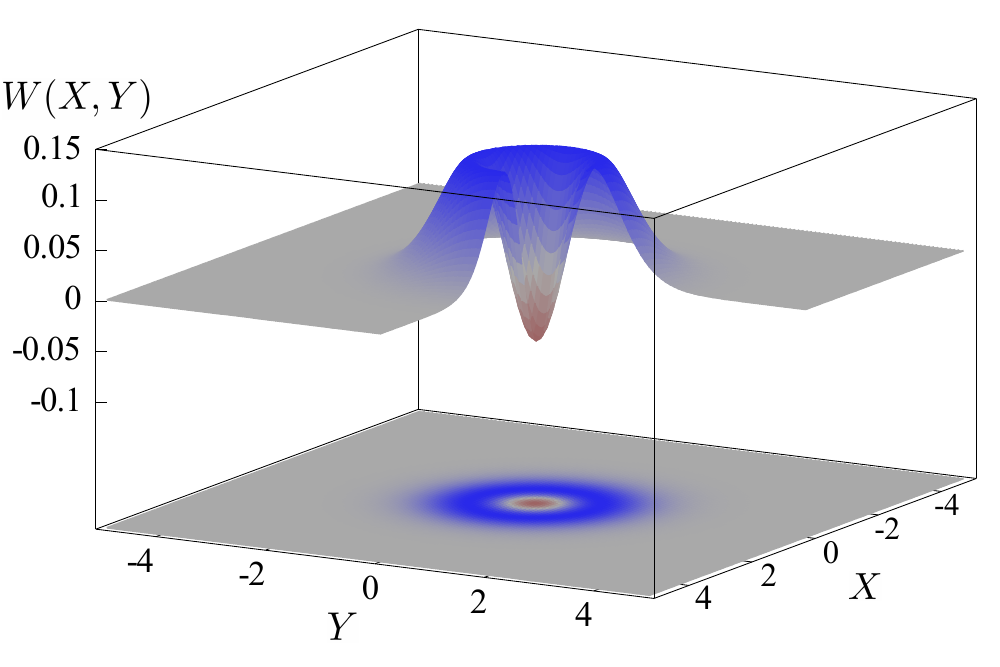}
\vspace{-3mm}
\caption{Reconstructed Wigner function of the single-photon subtracted, phase-randomized weakly squeezed vacuum state. 
The data is not corrected for any kind of optical loss. Around the origin, the Wigner function has negative values down to -0.063$\pm$0.004. This value achieves about 20\% of the strongest negativity possible of $-1/\pi \approx -0.318$, which corresponds to a perfect measurement on a pure single photon Fock state.}
\label{fig:Fock}
\end{figure}

The Wigner function of the density matrix $\tilde{\rho}$ can be expressed as
\begin{equation}
W(X,Y)= \frac{1}{\pi} \sum_{n=0}^N (-1)^n p_n L_n(2X^2+2Y^2)e^{-X^2-Y^2},
\end{equation}
where $L_n(x)$ denote the Laguerre polynomials. The Wigner function is plotted in Fig.~\ref{fig:Fock}. It clearly shows negative values around the origin, with a maximum negativity of $W(0,0)=-0.063\pm0.004$.
The statistical predicate on the negativity of the Wigner function was obtained via a bootstrap algorithm that was applied to the data and the reconstruction algorithm is repeated with the new data set.
The Wigner function is witnessed to be negative with more than 15 standard deviations.

\emph{Summary} --
Quantum state tomography of a phase-randomized nonclassical state at the telecommunication wavelength of 1550\,nm is performed using balanced homodyne detection. Without correction for any optical loss, the measurements yield a negative Wigner function with a value of $W(0,0)=-0.063\pm0.004$. The largest contribution to the state is a single-photon Fock state (57\%), whereas 39\% is contributed by the vacuum state. The state is generated by subtracting a single photon from a weakly squeezed vacuum state at a low transmission beam splitter. The tapped mode is detected with an avalanche photo detector with preceding frequency up-conversion to 532\,nm to allow for the use of room-temperature silicon detectors.
This work shows the detection of negative Wigner functions at the prominent wavelength of 1550\,nm by means of balanced homodyne detection and detection of heralding photons without devices cooled below room temperature.\\

The authors thank Sacha Kocsis, Mikhail Korobko and Axel Sch\"onbeck for fruitful discussions. 
This work was supported by the Deutsche Forschungsgemeinschaft (DFG) (SCHN~757/4-1, The Centre for Quantum Engineering and Space-Time Research, QUEST);
J.F. acknowledges financial support by the Czech Science Foundation (GB14-36681G).

%\bibliography{references}

\end{document}